\documentclass[12pt]{article}

\usepackage{scicite}
\usepackage{graphicx}
\graphicspath{ {./images/} }
\usepackage{times}
\usepackage{amsmath}

\newenvironment{sciabstract}{%
\begin{quote} \bf}
{\end{quote}}

\begin{document}


\title{Rapid imaging of special nuclear materials for nuclear non-proliferation and terrorism prevention}


\author
{Jana Petrovi\'c,$^{1}$ Alf G\"o\"ok,$^{1}$ Bo Cederwall$^{1\ast}$\\
\\
\normalsize{$^{1}$ Department of Physics, KTH Royal Institute of Technology, SE-10691 Stockholm, Sweden.}\\
\\
\normalsize{$^\ast$ Corresponding author: Email: bc@kth.se.}
}


\date{}




\baselineskip24pt


\maketitle


\begin{sciabstract}
We introduce a novel technique, neutron-gamma emission tomography (NGET), for rapid detection, 3D imaging, and characterization of special nuclear materials like weapons grade plutonium and uranium. The technique is adapted from fundamental nuclear physics research and  represents a paradigm shift in the approach to detection and imaging of small quantities of such materials. The method uses a granular detection system based on fast organic scintillators, measuring the characteristic fast time and energy correlations between particles emitted in nuclear fission processes. The radically new approach of using such correlations in real time in conjunction with modern machine learning techniques provides unprecedented imaging efficiency, spatial resolution and ultra-low false alarm rates, addressing global security threats from terrorism, the proliferation of nuclear weapons, as well as different nuclear accident scenarios and environmental radiological surveying.
\end{sciabstract}


\section*{Introduction}

While special nuclear materials (SNM) like plutonium and uranium form the basis for enormous technological developments in the last century they have also led to some of the most serious threats to humankind and its very existence. Even though the danger of a global military conflict involving nuclear weapons of mass destruction is at the center of public awareness in this context, the potential consequences from acts of terrorism involving such materials may not fall far behind. The stewardship and control of SNM is therefore of major importance for protecting the safety of present and future generations and it is one of the main concerns of international collaboration within the International Atomic Energy Agency (IAEA). Key to such efforts is the ability to efficiently and quickly detect, localize, and characterize SNM in different national security scenarios as well as in routine inspection of passengers, goods, vehicles, etc. at, e.g., sea- and airports and transport logistics hubs. This requires advanced sensor-based detection of threat objects, including radiation detection systems.
Among the most important materials of interest are weapons grade uranium (WGU), weapons grade plutonium (WGP) and radioactive sources that could be used for radiological dispersion devices (RDD). An effective defense against these and similar threats must take advantage of the latest developments in radiation detection technology. Much-needed advances in this field include not only enhanced sensitivity of detection but also the possibility to perform efficient localization of SNM in screening operations of, e.g.,  objects, persons and vehicles in the field.  

The detection of SNM in Radiation Portal Monitors (RPMs) is one of the critical links in the nuclear security chain. RPMs are designed to alarm if the measured radiation flux, primarily of neutrons and $\gamma$-rays, exceeds predefined thresholds when persons, vehicles, packages or other objects are passing through them. Once a source is detected, it would be highly beneficial to rapidly discern its precise  location. Current RPM systems do not have such capabilities. Nuclear emergency responders and safeguards inspectors would also benefit significantly from the ability to rapidly verify the location of a source without close manual intervention.

Since the presence of neutrons is a key indicator of materials undergoing spontaneous or induced fission, neutron detectors are crucial for the detection of SNM. Thermal neutron gaseous proportional counters based on $^3$He are currently the most common detectors used in nuclear safeguards and nuclear security systems, such as in RPMs \cite{Ensslin1998,Menlove1985}. However, due to the global shortage of $^3$He there is an increasing interest in developing $^3$He free systems \cite{Pickrell2013}.
In addition to proportional counter systems replacing $^3$He with other high thermal-neutron absorption cross-section materials there is also an increasing focus on fast-neutron detection based on organic scintillators.
While thermal neutrons typically do not carry any useful information of their initial energy or emission point, fast-neutron detection opens new possibilities for imaging SNM.
Several types of neutron imagers have been developed, including coded-aperture imagers \cite{Ress1988,Hausladen2012}, time-encoded imagers \cite{Marleau2011}, and neutron scatter cameras \cite{Mascarenhas2006,Ryan2008,Gerling2014,Tagawa2019,Madden2013,Poitrasson2014,Steinberger2020}. While neutron scatter cameras can be made more compact than time-encoded or coded aperture  neutron imagers their efficiency suffers greatly from the requirement of detecting two consecutive neutron scatters.  Furthermore, a fundamental complication in the design of neutron scatter cameras is the need to maximize the probability that the incident neutron scatters only once in the scatter detector and subsequently escapes so that it may interact in another detector element. If the scatter detector is too thick, additional scatters become likely, corrupting the energy measurement and making it impossible to reconstruct the neutron’s incident direction. Attempts to circumvent this problem would lead to reduced detection efficiency or to complex and costly detection systems.

Even though the main focus of previous studies has been on neutron detection there are potential advantages associated with the additional detection of $\gamma$-rays as a means to identify the presence of special nuclear materials with very high sensitivity. These are mainly the significantly higher multiplicity of prompt fission $\gamma$-rays and their short flight time from source to detector. Most of the $\gamma$-rays in fission are emitted in prompt cascades depopulating short-lived (typically picoseconds or shorter) excited states in the fission products, with a multiplicity distribution that can extend significantly beyond an average of 5–10 \cite{Lemaire2005,Hamilton1995,Bleuel2010}. Therefore, in particular fast $\gamma$-neutron correlations deserve further attention in the development of more sensitive methods for nuclear safeguards and security applications~\cite{Trombetta2019}. Such techniques have been used in fundamental nuclear physics experiments for decades, see Ref.~\cite{Cederwall2020} for a recent example.  An additional advantage of fast $\gamma$-neutron coincidence detection compared with detection of neutrons alone is the energy independence of the photon time of flight, which reduces the correlation time spread between the detected particles. It was previously proposed to use time-of-flight information for event-by-event 3D imaging of SNM in neutron scatter cameras \cite{Monterial2017} by combining the fast coincidence detection of a neutron scattering between two detector elements with the detection of a correlated photon in a third detector element. However, for such triple-coincidence systems the detection efficiency will be significantly reduced compared with the already low efficiency of neutron scatter imaging and it may not be useful for applications where a rapid assessment is needed.

The velocity distributions of neutrons emitted in the fission processes are well characterized for the most common nuclides present in SNM. This information is used in the present work in order to deduce the position of a source from measured fast $\gamma$-neutron time-of-flight correlations.

We here present a new approach, neutron-gamma emission tomography (NGET), to 3D localization of SNM using fast $\gamma$-neutron coincidence detection applicable to national security measures, nuclear emergency response, nuclear safeguards, environmental radiological surveying and related fields. Two techniques are presented, both exploiting the properties of correlated $\gamma$-neutron pairs originating from the same fission events. 
We first show results from an implementation of this approach based on estimating a region of origin from the measured energy deposited by the neutron and the $\gamma$-neutron time difference, event by event.
Since only the detection of a single neutron interaction is required, the neutron energy is \emph{estimated} based on sampling the approximately known kinetic energy distribution of fission neutrons above the detected recoil energy, which is the minimum kinetic energy that could be carried by the incident neutron. 
The method has similarities with emission tomographic methods used in medical imaging, such as positron emission tomography (PET)~\cite{Ter-Pogossian1975,Phelps1975}. PET imaging uses the fact that 511~keV photon pairs from positron annihilation are strongly correlated, directionally in space and in time, in order to deduce the distribution of positron-emitting isotopes within the field of view. A PET detector system with good enough ($\sim$ sub-nanosecond) time resolution can, additionally, exploit the relative time-of-flight (TOF) information for the detected photon pairs to improve on its 3D image resolution~\cite{Ter-Pogossian1981}.
Differently from positron annihilation, the physics of the nuclear fission process does not provide easily deduced direct directional correlations between the emitted photons and neutrons. Neutrons, being massive particles, have a velocity that is directly related to their kinetic energy. Therefore, since neutrons from spontaneous fission of SNM have rather well established energy distributions, often approximated by a Watt spectrum~\cite{Watt1952} with parameters depending weakly on the nuclide, it is possible to estimate the probability that a detected neutron had a certain incident velocity  based only on a partial energy measurement or without measuring the energy at all. Using such estimates, the measured relative time-of-flight between the neutron and the photon in a correlated neutron-photon pair can then be translated into information about their point of origin.
Similarly to a neutron scatter camera this technique then relies on standard tomographic image reconstruction techniques. We here present results obtained using a de-convolution algorithm based on Bayes' theorem \cite{DAgostini1994}.

In another implementation of NGET imaging we rely on \emph{cumulative} distributions of correlated $\gamma$-neutron time differences recorded between different pairs of radiation sensors for rapid, on-line, reconstruction of the source position. A particular advantage of this cumulative NGET (CNGET) technique is that it only requires the difference in detection time between correlated particles. Hence, it is enough that the detector elements have a fast (nanosecond) time response and that they are sensitive to both gamma rays and neutrons. The CNGET technique therefore does not, a priori, require the usually more expensive ability to discriminate between neutrons and $\gamma$-rays, nor does it require measurement of particle energies. This technique relies on machine learning algorithms or statistical iterative methods to accurately localize SNM. 

By localizing unknown sources in 3D, both implementations of the NGET imaging method also provide the basis for identifying, quantifying and characterizing small amounts of SNM.

\section*{Results}
In this work, rapid and accurate localization of SNM samples with NGET is demonstrated using an RPM prototype system developed at the Royal Institute of Technology (KTH)~\cite{Trombetta2020}. A basic requirement on the detection system is an array of radiation sensors with fast time response, sensitive to fast neutrons and $\gamma$-rays. Most favorably the sensors are simultaneously sensitive to both fast neutrons and $\gamma$ rays and have a high capability to discriminate between the two types of particles. This is the case for the organic scintillation detectors employed in the present work, which provide efficient $\gamma$-neutron discrimination and fast timing, of the order of 1 ns. 
The RPM prototype system consisted of an array of eight 127 mm diameter by 127 mm length cylindrical liquid organic scintillator cells arranged in two detector assemblies. Measurements have been carried out using a californium-252 (Cf-252) radioactive source with mass $3.2\times10^{-9}$~g encapsulated in a 4,6 mm diameter x 6 mm cylindrical ceramic casing. Cf-252 has spontaneous fission as a 3\% decay branch (the remaining 97\% being due to $\alpha$ decay) and an average prompt-neutron multiplicity of 3.76 per fission~\cite{Croft2020}. The source is equivalent to around 100 g of weapons grade plutonium (7\% plutonium-240 and 93\% plutonium-240) in terms of its approximately 1900 s$^{-1}$ fission rate. This would correspond to a sphere of radius approximately 1 cm metallic plutonium. Details of the setup and the measurements are described in the Materials and Methods section provided in the supplementary materials.

\subsection*{NGET imaging performance}
\begin{figure}[ht]
\centering
\includegraphics[width=0.6\columnwidth,clip]{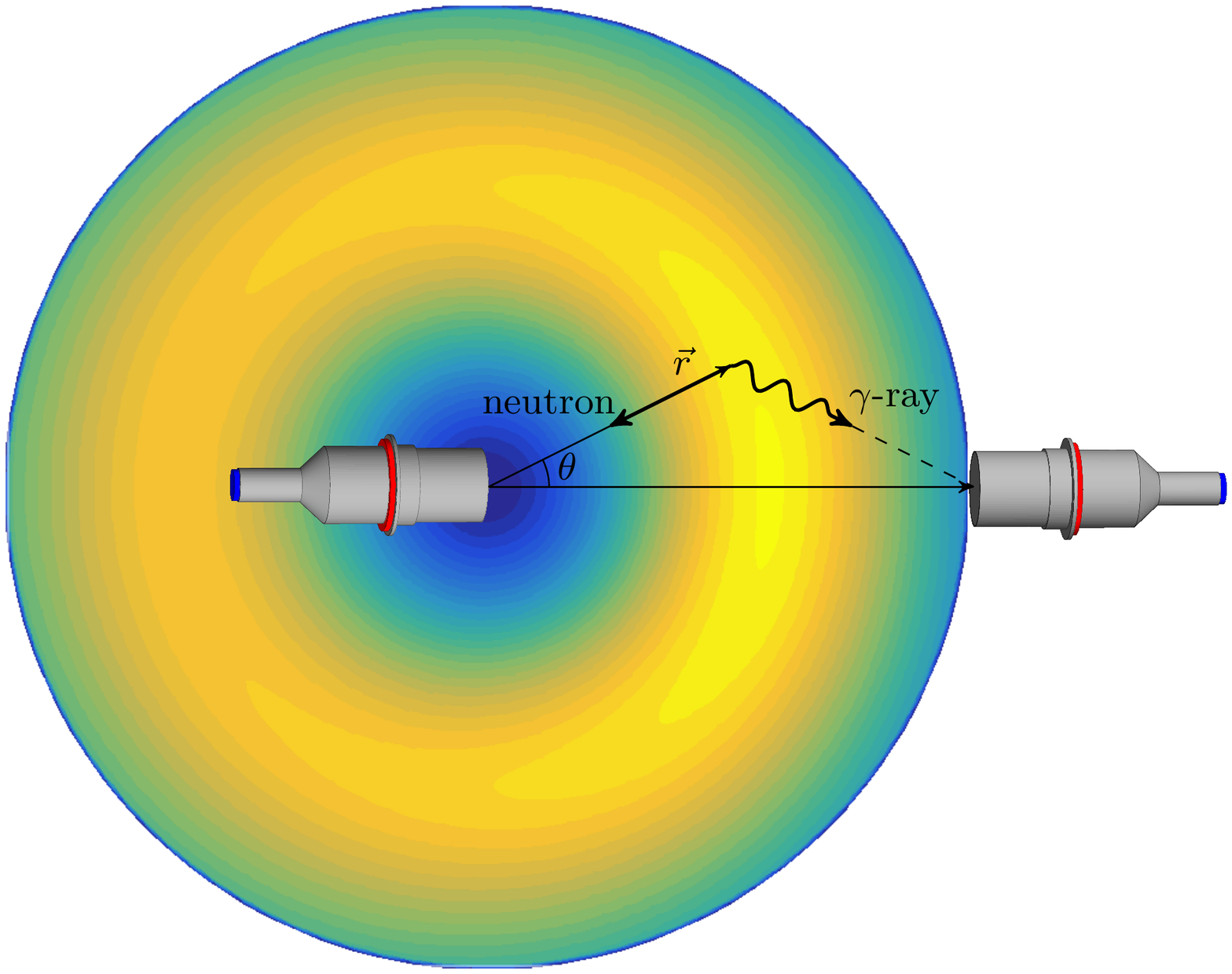}
\caption{Schematic illustration of NGET event mapping following the detection of a correlated $\gamma$-neutron pair from one fission event. The probability that the fission event occurred at the position $\vec{r}$ relative to the center of the detector registering the neutron is indicated as a function of the angle and distance to the source position by an arbitrary color scale for a given time difference between neutron and $\gamma$-ray detection.}
\label{fig:Bayes_illustration} 
\end{figure}
We first present results from applying the novel NGET imaging principle with the KTH RPM prototype using the technique based on event-by-event mapping of possible points of origin using probability density functions (PDF). The PDFs are derived from the measured neutron recoil energy and relative detection time of $\gamma$-neutron pairs which are correlated from the same fission event. The principle is illustrated schematically in Figure~\ref{fig:Bayes_illustration}. For each correlated detection of a neutron and a $\gamma$-ray in two specific detector elements the neutron and $\gamma$ flight paths are well defined as a function of the possible  position of a fission event in space. Knowledge about the prompt fission neutron energy spectrum shape can then be used to extract the probability distribution for the unknown source position given the observation of a certain time difference between $\gamma$-ray and neutron detection. This probability distribution can be further constrained using information on the detected neutron recoil energy. 
The measured neutron-recoil energy is in most cases due to one or more elastic scatterings on protons in the detector medium. Neutrons and protons have approximately the same mass so the scattering process is similar to classical billiards. For each energy deposition in the detector material from a recoiling proton, the incident neutron energy must be higher or equal to this value.
The detector energy response to a single neutron interaction is an approximately known non-linear function of the deposited energy. The full detector response as a function of incident neutron energy, which is highly dependent on the incident neutron energy and the geometrical dimensions of the detector, can also be calculated approximately or measured.
Hence, the approximately known energy distribution of fission neutrons can, in combination with the detector response function, be used to estimate the probability distribution for the incident neutron energy. 
Each $\gamma$-neutron coincidence event hence gives rise to a unique PDF, depending on which detector elements were involved, the measured time difference between the $\gamma$-ray and neutron interactions, and the detected neutron energy. By combining the information from several such events a combined PDF for the source location can be deduced. In this way, the uncertainty of the source location is successively reduced for each correlated $\gamma$-neutron event that is detected. 

\begin{figure}[hptb]
    \centering
    \includegraphics[width=\textwidth]{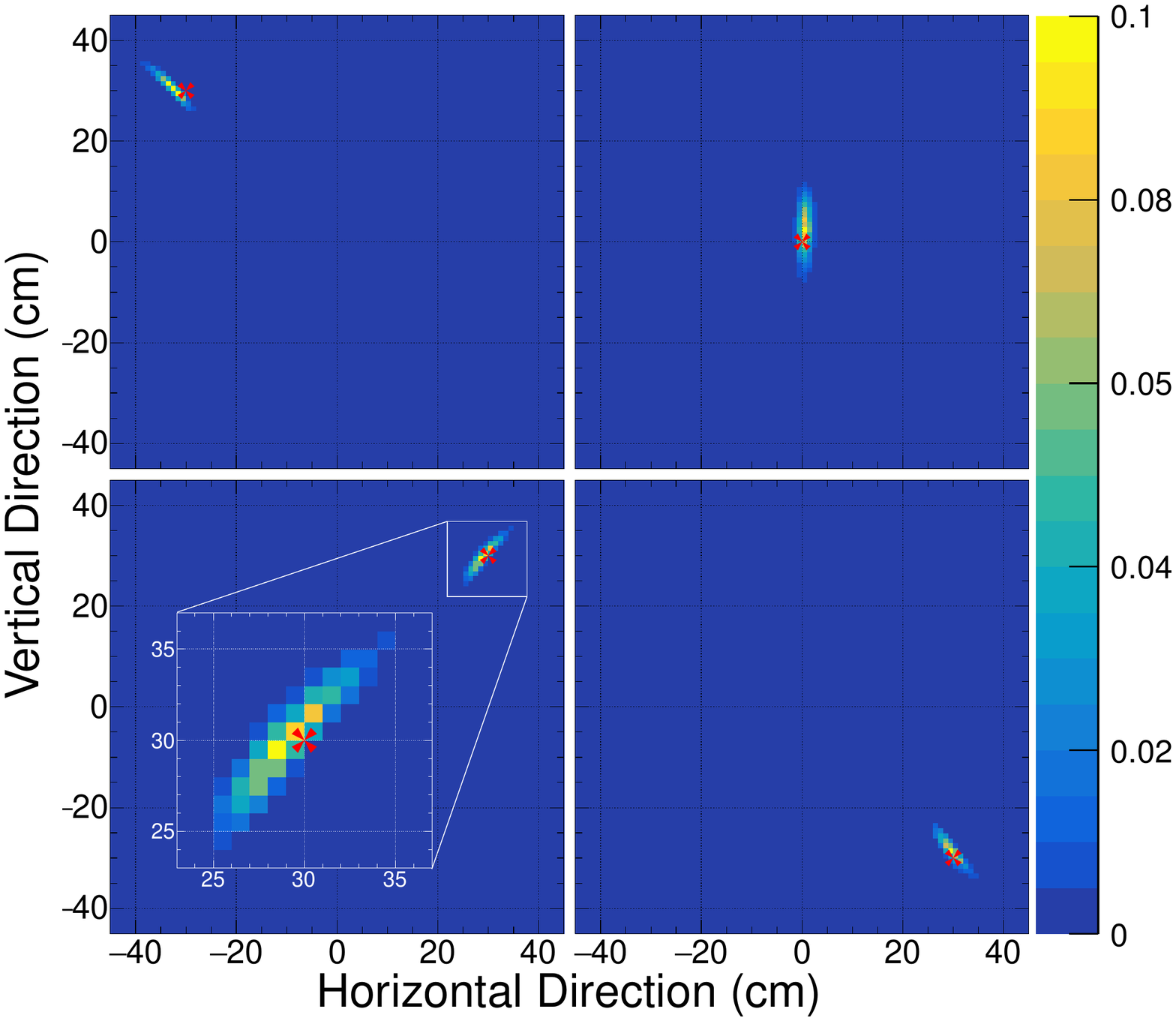}
    \caption{Demonstration of the source localization obtained using the Bayesian inference algorithm. Results of 10-second measurements with the 1.7~$\mu$Ci $^{252}$Cf source described in the text placed in four different positions are shown as projected images in the horizontal-vertical plane. The estimated probability per cm$^2$ of finding the source as a function of position is indicated by the color scale. The actual positions of the source during the measurements are marked with red crosses. In the lower left panel, a magnified view around one of the positions is shown in the inset.}
    \label{fig:Bayes_localization} 
\end{figure}

The algorithm developed for this purpose uses Bayesian inference, and is described in detail in the supplementary material. The result of the algorithm is the PDF of finding the source as a function of the position. The application of the algorithm to experimental data from the KTH RPM prototype system with the source in different positions is illustrated in Figure~\ref{fig:Bayes_localization}. After 10 s of measurement the source location PDF has a standard deviation of about 4 cm. A longer measurement time of 30 s, reduces the standard deviation by about a factor of 2. For longer measurement times the point spread function that is achieved with this method for the current detector configuration is 1 cm.

\begin{figure}[h]
    \centering
    \includegraphics[width=0.5\textwidth]{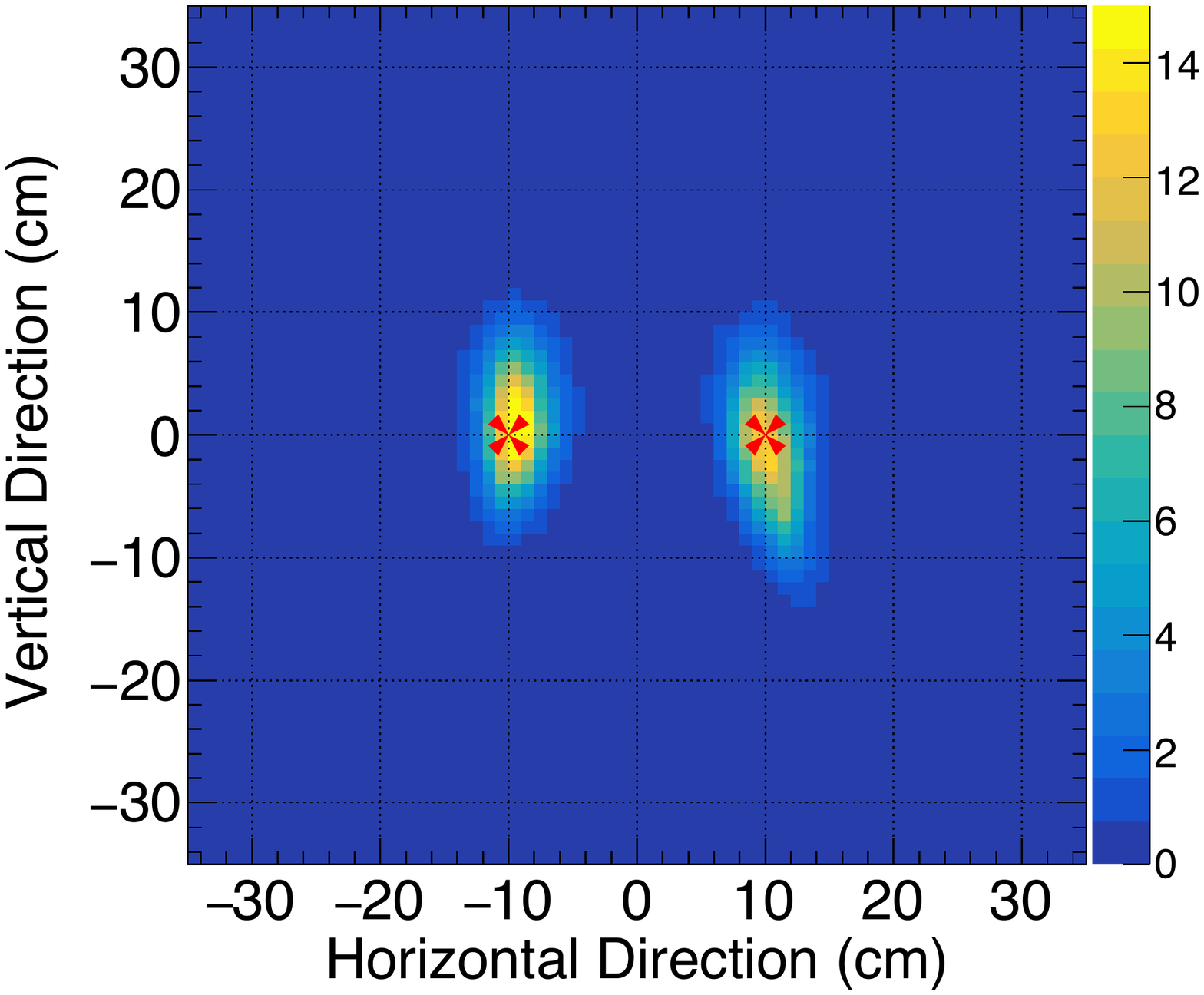}%
    \caption{Projected image in the vertical-horizontal plane obtained for two Cf-252 sources separated by 20 cm and measured during 1 min.
    The activity of the source on the right (left) is 1.7$\mu$Ci (0.56$\mu$Ci) corresponding to 1900 (630) fissions/s. The actual positions of the sources are marked with red crosses. The color scale indicates the number of detected events per cm$^2$.}
    \label{fig:Bayes_localization_two_sources}
\end{figure}

The Bayesian inference algorithm for source localization assumes that all detected events originate from the same point and it can  therefore not simultaneously identify the positions of multiple sources. For this purpose a different algorithm is applied, which has a somewhat lower spatial resolution for a single source but is capable of imaging multiple sources and/or an extended source distribution. The application of this algorithm to the measurement of two
Cf-252 sources, one with approximately three times lower activity, is shown in Figure~\ref{fig:Bayes_localization_two_sources}.

\subsection*{NGET imaging based on cumulative time difference distributions (CNGET)}
\begin{figure}[h!]
    \centering
    \includegraphics[width=0.9\textwidth]{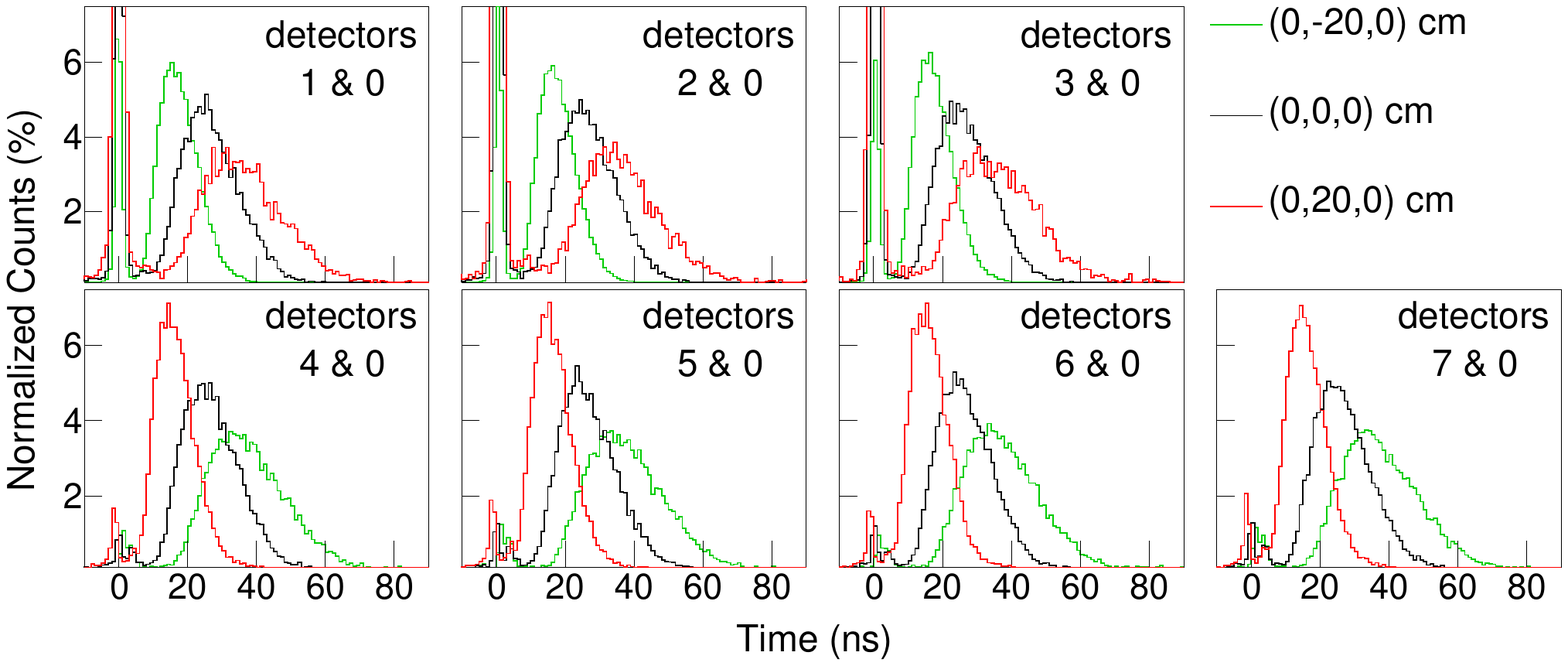}
    \caption{Examples of measured $\gamma$-neutron time difference distributions for three different source positions relative to the center of the detector system (see supplementary materials for details): green:(x,y,z)=(0,-20,0) cm, black:(x,y,z)=(0,0,0) cm and red: (x,y,z)=(0,20,0) cm. Each panel corresponds to the measured time differences between the detection of a neutron in a specific sensor position (see panel label) and the correlated detection of a $\gamma$-ray in sensor position \#0. The time differences are limited to the range [-10,90] ns which is the useful correlation time for the present RPM geometry}
    \label{fig:ng_tdd}
\end{figure}

CNGET 3D imaging is a simplified approach to full NGET providing rapid and accurate 3D-localization of SNM based solely on measuring the accumulated $\gamma$-neutron arrival time difference spectra for different combinations of detector elements. In the present detector system there are eight different sensor elements and therefore $8 \times 7=56$ unique combinations of $\gamma$-neutron time difference distributions. Assuming a point source of time correlated neutrons and $\gamma$ rays such as a small amount of SNM, each time distribution contains unique information regarding its position. For many detection geometries, such as the present one, it may be assumed that the photon is detected before the neutron since the mean velocity of neutrons from nuclear fission is roughly an order of magnitude smaller than the speed of light. Therefore, the method does not require $\gamma$-neutron discrimination capabilities, nor does it require measuring the particle energies.
The CNGET technique employs a set of time difference distributions, each corresponding to one unique pair of detector elements, which are accumulated continuously during the measurement time. The measurement time will vary depending on the application, from short (seconds - minutes) time scales in the case of security and emergency scenarios to longer time scales, e.g. for spent fuel verification in nuclear safeguards and for environmental surveying. Figure~\ref{fig:ng_tdd}  shows examples of $\gamma$-neutron time difference distributions obtained from measuring on the same Cf-252 radioactive source as described above, placed in different positions. As can be seen in the figure, these distributions have different characteristics depending on the position of the source. Hence, they can be used to provide the location of the source in space with the aid of, e.g., machine learning. 

\begin{figure}[h!]
    \centering
    \includegraphics[width=0.7\textwidth]{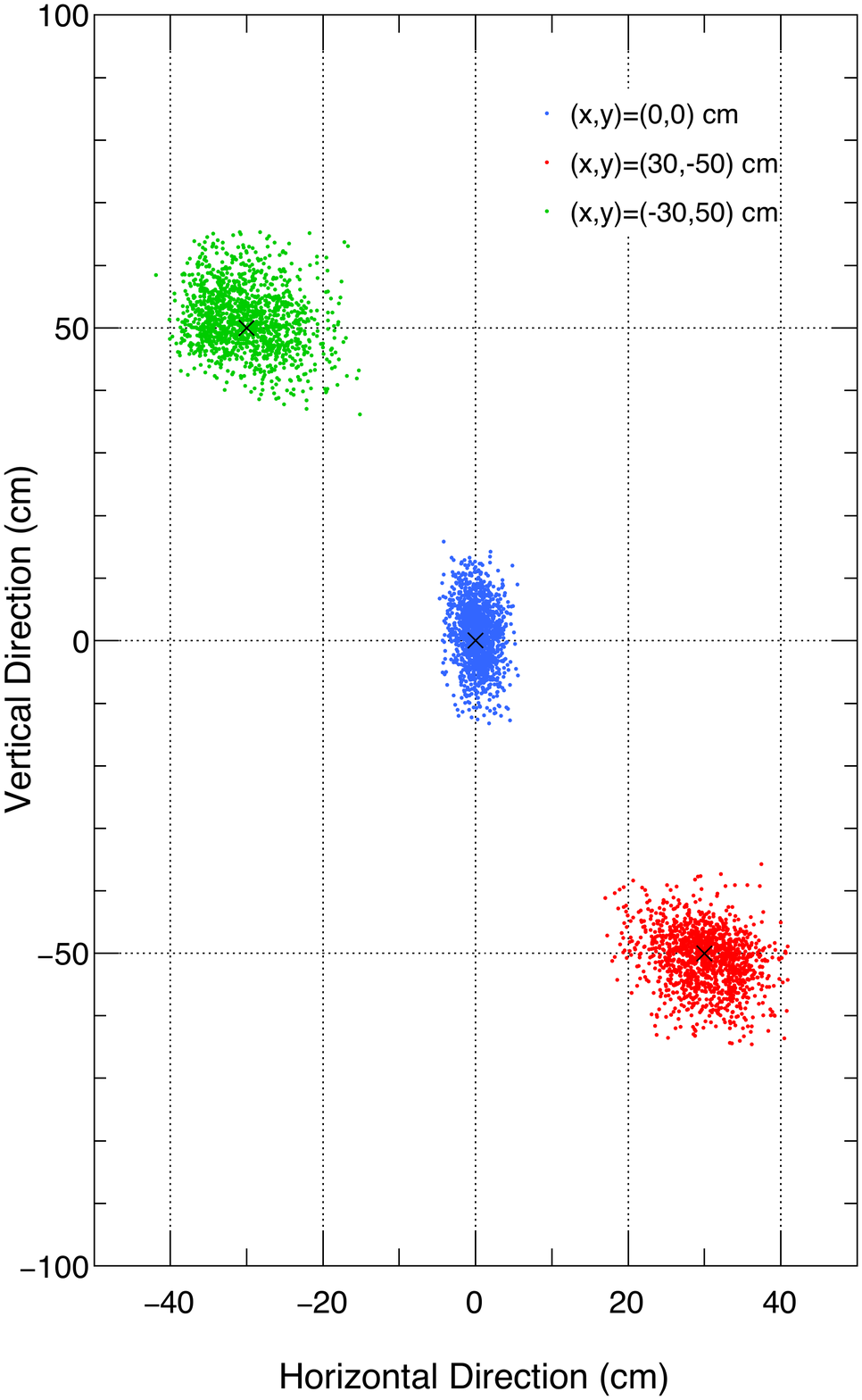}
    \caption{CNGET localization results for the Cf-252 source described in the text. The results are shown for three different source positions, as indicated in the legend, within the central plane of the detector system. Each point corresponds to one minute measurement time.}
        \label{fig:ng_scatterplot}
\end{figure}

In the present study we have trained an artificial neural network (ANN) algorithm using the deep learning toolbox in the MATLAB$^{TM}$\cite{MATLAB:R2020a_u4} computational software package. Figure~\ref{fig:ng_scatterplot} shows the results obtained for the Cf-252 source described above when placed in three different positions, each point measured during one minute. 

\section*{Discussion}
In order to ensure public safety and international security, the risk that nuclear or other radioactive materials could be subject to illegal trafficking and used in criminal or other unauthorized acts must be minimized.  This requires versatile detection systems that can rapidly detect SNM and other radiological threats in real time. The possibility to rapidly and precisely locate SNM, e.g., at border controls, postal mail centers or transport hubs  would dramatically increase the effectiveness of counter-terrorism measures and non-proliferation efforts. 
In the recent work of Steinberger et al.~\cite{Steinberger2020} a state-of-the-art compact 2D radiation imaging system for SNM was presented based on the principle of neutron scatter imaging with organic scintillation detectors. The here presented NGET and CNGET techniques enable localization of SNM in 3D with higher accuracy and orders of magnitude higher efficiency. Although there is a significant difference between the detector system used in the present work and that employed by Steinberger et al. in terms of detection geometry, in the total volume of the active detector medium, etc., an important reason for the higher performance obtained here is due to the use of the fundamental properties of $\gamma$-fast neutron correlations from nuclear fission, in particular the high $\gamma$-ray multiplicity~\cite{Trombetta2020}. 
Both NGET techniques presented here can be used to rapidly localize small amounts of SNM with high accuracy. This promises a paradigm shift in national security applications, nuclear safeguards, environmental surveying, and for nuclear emergency responders. In addition, the cumulative NGET (CNGET) technique only relies on the fast timing properties of the sensor elements and can therefore be implemented with relatively modest technical modifications into standardized RPM designs using conventional plastic scintillators.
 On the other hand, the image resolution improves more rapidly as a function of measurement statistics (i.e. measurement time) with full event-by-event NGET imaging due to the additional information on the neutron kinetic energy that is included in the image reconstruction. An additional advantage is that the event mapping technique employs an analytical probabilistic approach to the reconstruction of the source location as opposed to empirical machine learning methods. 

It is noteworthy that the RPM detector system employed in the present work has not been optimized for efficiency nor for imaging purposes. Significantly improved spatial resolution and response uniformity could be obtained by using a set of more uniformly distributed detectors or adopting a higher detector granularity (i.e. smaller detector cells). Still, the results are remarkably good with an average position uncertainty of 4.2 cm for a 1-minute measurement using this relatively weak source (approximately 35$\%$ of the ANSI N42.35-2016 industry standard~\cite{ANSI-N42.35-2016} benchmark Cf-252 source for RPM characterization).


\bibliography{scibib}

\bibliographystyle{Science}

\section*{Acknowledgments}
This work has received funding from the Swedish Radiation Safety Authority under contract Nos SSM2016-3954, SSM2018-4393, and SSM2018-4972. Raw data are available upon request to the corresponding author.
\section*{Supplementary materials}
\section*{Materials and Methods}

\subsection*{Detection system}
\begin{figure}[h!]
    \centering
    \includegraphics[width=0.5\textwidth, clip, trim=70 130 0 50]{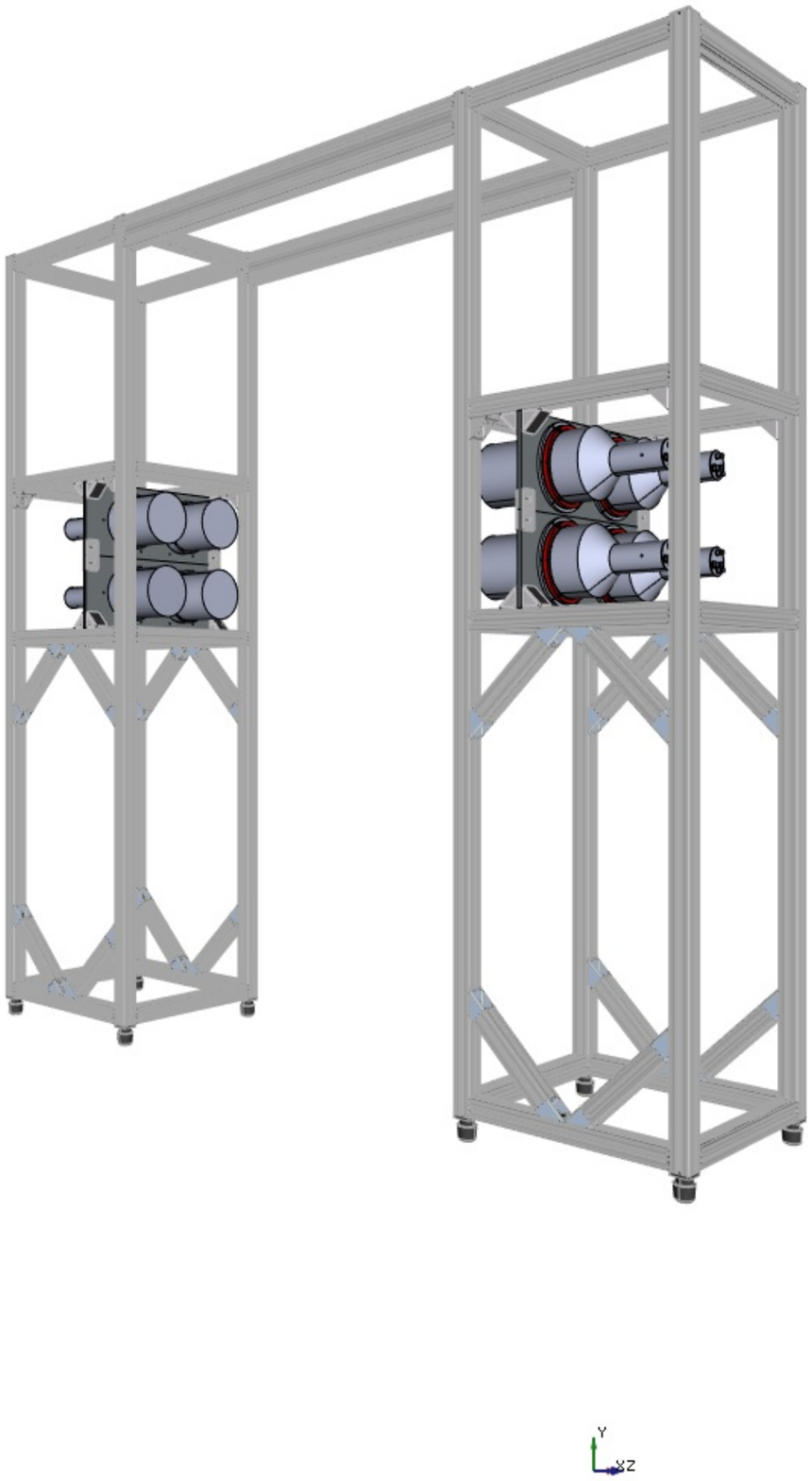}
    \caption{CAD model of the detector system including its mechanical support structure.}
        \label{fig:cad_drawing}
\end{figure}

The 3D imaging capabilities of NGET are developed and illustrated within the framework of an RPM system. However, the method is applicable to any radiation detector system containing a minimum of two geometrically displaced sensor elements with a fast (ns) time response and with at least one detector element sensitive to fast neutrons and at least one sensor element sensitive to $\gamma$-rays. Some organic scintillators are well suited for this purpose as they are sensitive to and may discriminate between fast neutrons and $\gamma$-rays using pulse shape analysis (PSA). They also have a fast time response (typically around 1 ns or better) as well as limited spectroscopic capabilities. The RPM prototype system constructed at KTH~\cite{Trombetta2020} that we use for illustrating the new imaging method consists of an array of sensors placed in two vertical pillars with a geometry as specified in the ANSI N42.35-2016 industry standard \cite{ANSI-N42.35-2016}, together with the associated electronics.  In the current configuration a total of eight 127 mm diameter by 127 mm length cylindrical EJ-309 scintillator \cite{Eljen} cells are arranged in two two-by-two detection assemblies, one in each pillar. The EJ-309 scintillator was chosen for its excellent $\gamma$-neutron discrimination properties~\cite{Kaplan2013} which enabled the RPM to detect single neutrons with a low level of contamination from background $\gamma$-rays \cite{Trombetta2020}. $\gamma$-neutron discrimination is not strictly needed for NGET since correlated neutrons and $\gamma$-rays are typically well separated by their relative detection times. However, the results are somewhat improved by measuring the neutron recoil energy in event-by-event mode. A CAD model of the detector system including its mechanical support structure is shown in Figure~\ref{fig:cad_drawing}. 

It is worthwhile to point out that the imaging capabilities of the system are not dependent on the object being inside the portal, or that the system contains two separate pillars of sensor elements. Each detection assembly has its own inherent imaging capabilities and different, more compact, sensor geometries are also possible, providing “single view” imaging.

Although other organic scintillator materials with excellent properties like stilbene \cite{Zaitseva2018} are available the EJ309 scintillator was chosen as detection medium as it features a number of chemical properties recommending it for use in environmentally challenging conditions outside the lab. These properties include a high flash point, low vapor pressure, and low chemical toxicity~\cite{Eljen}. Each detector cell is optically coupled to a Hamamatsu R1250 photomultiplier tube. The four photomultiplier tubes in each detection assembly are powered by a 4 channel CAEN DT5533 high voltage power supply. The photomultiplier tube anode pulses are acquired and digitized with an eight channel CAEN DT5730 digitizer board featuring a 2Vpp dynamic range, 14 bit resolution, and 500 MHz sampling rate. The time synchronized data is transferred using a USB 2.0 link to a personal computer (PC) running on a Linux or Windows™ platform. Besides the raw data transfer of detector waveforms the digitizer also features firmware programmable pulse shape analysis (PSA) using two field programmable gate arrays (FPGA) of type Intel/Altera Cyclone EP4CE30, each handling four of the ADC data streams. The digitized signals from the RPM detector modules can be processed in real time within the digitizer’s FPGAs to extract charge integrals, pulse shape information and timing information. The data are transmitted via USB to an industrial PC, where further processing is performed. The PSA algorithm for distinguishing $\gamma$ rays from neutrons is based on the charge comparison method \cite{Wolski1995} whereas the time-of-arrival information is extracted using a digital constant fraction discrimination algorithm.  Energy information for each trace is deduced using a moving window de-convolution algorithm. The processed data stream contains information on the individual detected neutrons and $\gamma$-rays (energy, time, and the sensor element that registered the hit) as well as time averaged single-neutron and $\gamma$ rates, fast coincidence rates for $\gamma$-neutron, neutron-neutron, and $\gamma$-$\gamma$ events with adjustable coincidence time windows and thresholds. Typical coincidence time windows are 0-100 ns for $\gamma$-$\gamma$ and neutron-neutron events and 10-100 ns for $\gamma$-neutron events.

\subsection*{Methods}

\subsubsection*{Source localization using Bayesian Inference}
The NGET algorithm to find the location of a fission source is based on Bayes' theorem, which gives a way to estimate the probability of a hypothesis $H$ to be true, given the occurrence of an event $E$ 
\begin{equation}
P(H | E) = \frac{P(E | H) P(H)}{P(E)},
\end{equation}
where $P(E | H)$ is the probability of the event to occur given that the hypothesis is true, $P(H)$ is the probability that the hypothesis is true and $P(E)$ is the total probability that the event occurs.

\begin{figure}[ht]
\centering
\includegraphics[width=0.8\columnwidth,clip]{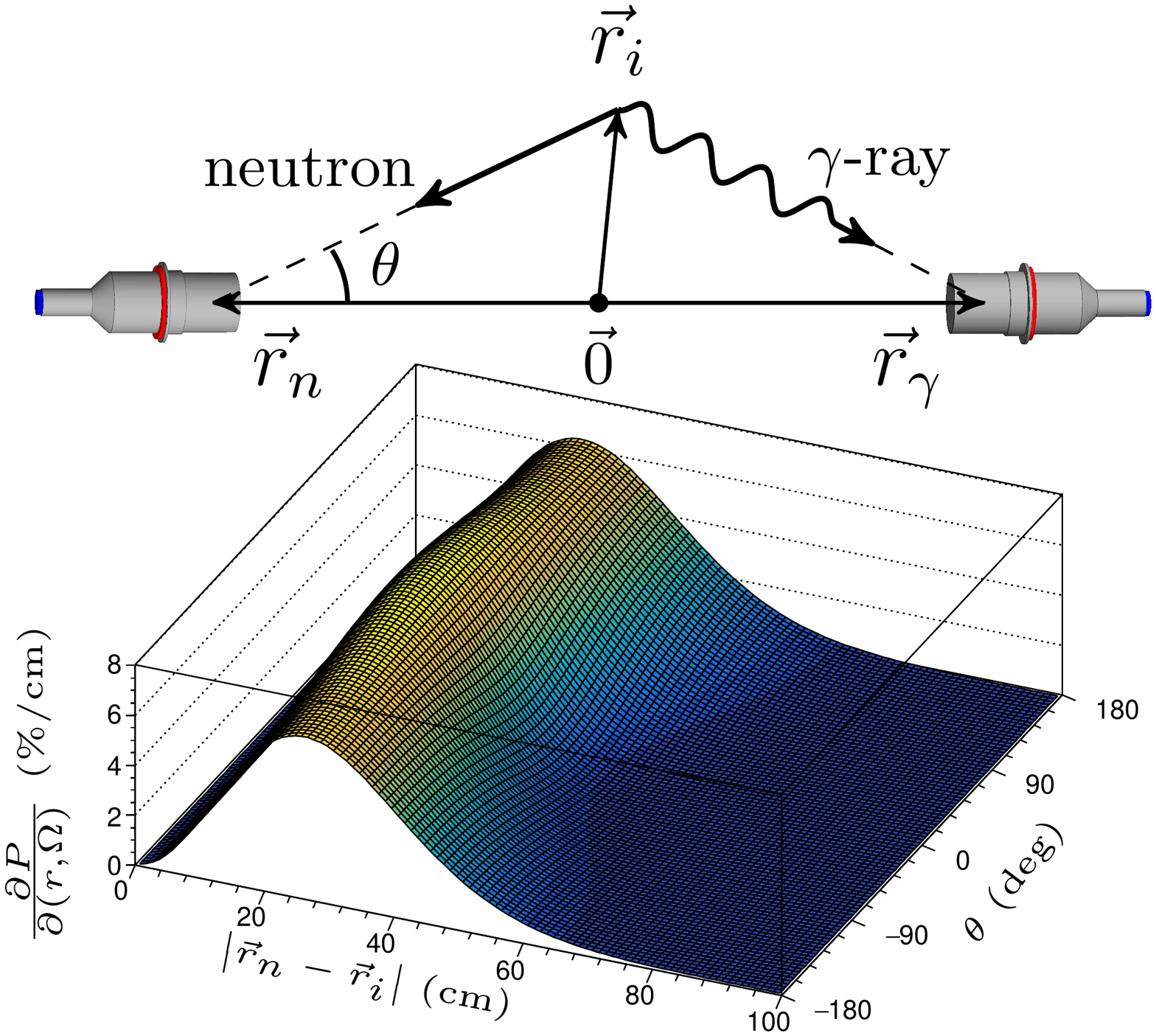}
\caption{Illustration of the Bayesian source-localisation algorithm. The probability to find the source at the position $\vec{r}_i - \vec{r}_n$, calculated for a fixed time difference between neutron and $\gamma$-ray detection $\Delta t_{n \gamma}=10$~ns, is shown as a function of the angle and distance to the source position as seen from the neutron detector.}
\label{fig-Bayes-illustration-2}       
\end{figure}

We define the event $E$ as the detection of a $\gamma$-neutron coincidence, with a time difference $\Delta t_{n \gamma}$ between $\gamma$-ray and neutron detection and a detected neutron-recoil energy, in terms of the light output $L$. For each voxel in space we form the hypothesis $H_{i}$, that the event originated from the center of the voxel $\vec{r}_i$. The time difference can then be written
\begin{equation}
\Delta t_{n \gamma} = t_{n} - t_{\gamma} =
\frac{\left|\vec{r}_n - \vec{r}_i \right|}{v_{n}} - \frac{\left|\vec{r}_{\gamma} - \vec{r}_i \right|}{c},
\end{equation}
where $\vec{r}_n$ and $\vec{r}_{\gamma}$ are the positions of the detector elements hit by the neutron and $\gamma$-ray, respectively. For a given position $\vec{r}_i$ the neutron time-of-flight is then

\begin{equation}
t_{n} = \Delta t_{n \gamma} + \frac{\left|\vec{r}_{\gamma} - \vec{r}_i \right|}{c}.
\end{equation}
From the neutron time-of-flight its kinetic energy $E_{n}$ is calculated. The probability that a neutron with this kinetic energy is emitted is taken from evaluated prompt fission neutron spectral data \cite{ENDF-B-VIII.0}. The probability density function (PDF) of possible event origins forms a fuzzy semispherical shell around the neutron detector position in space, see the illustration in Fig.~\ref{fig-Bayes-illustration-2}. 

The predominant interaction of neutrons in the organic scintillator is due to elastic scattering off protons. The neutron mean free path in the detector medium is on the order of a few centimeters for typical fission neutron energies. Therefore, the probability to have more than one scatter within a detector element can be relatively large, even for modest detection volumes. Each scattering interaction transfers a portion of the neutron's kinetic energy to the recoiling proton. A fraction of this energy is converted into fluorescent light in the scintillator. In the measurements, the scintillation light is converted to an electrical current pulse in the photomultiplier tube. The charge contained in this pulse is then integrated in order to provide a measure of the deposited energy.

\begin{figure}[ht]
\centering
\includegraphics[width=0.8\columnwidth,clip]{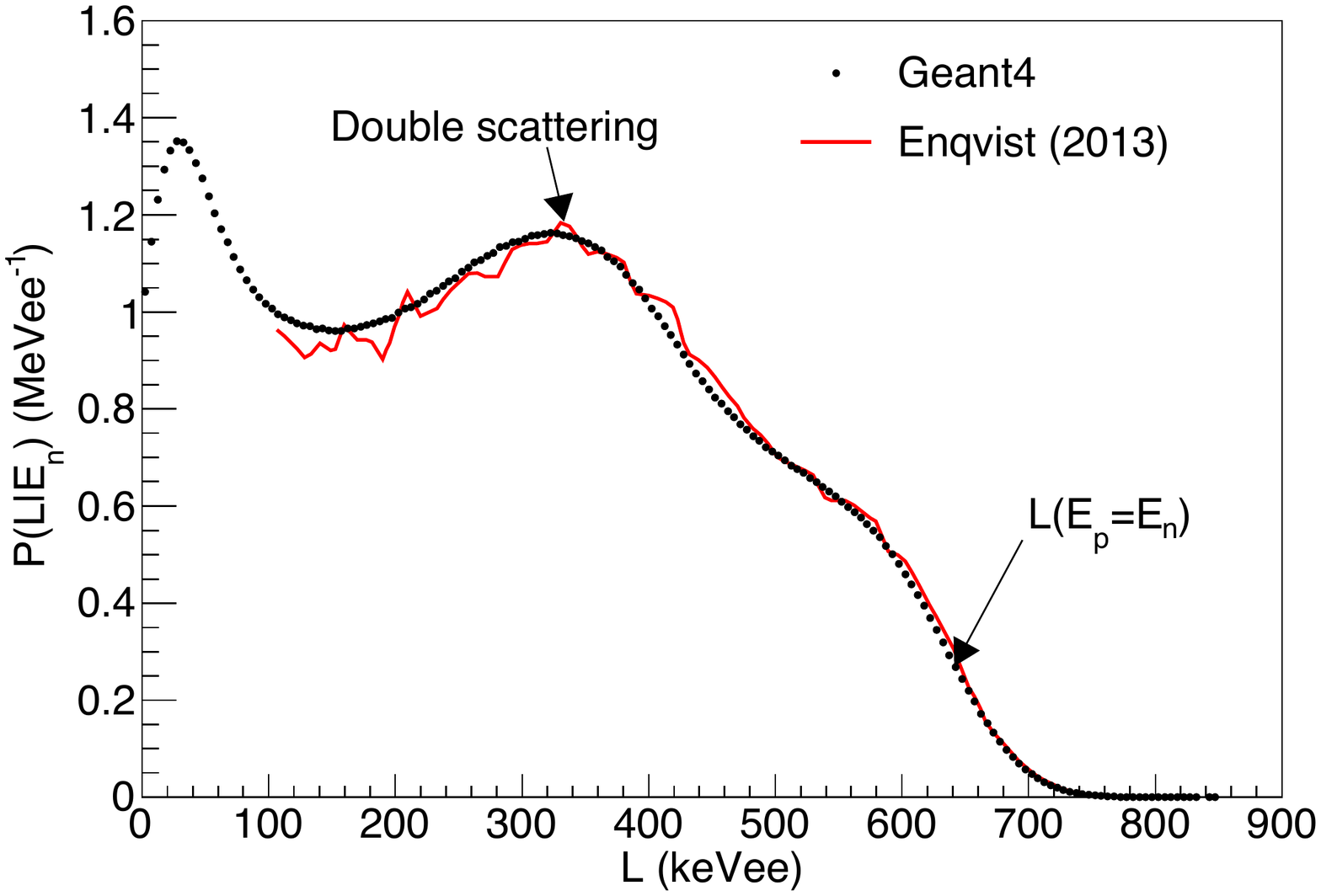}
\caption{Example of the light-output response function $R(L|E_n)$ for $E_{n}=2.45$~MeV. The Geant4 calculation is shown as black dots, while the red line represent experimental data from Enqvist et al. \cite{Enqvist2013}. The mean light output for protons with the maximum kinetic energy $E_{p}=E_{n}$ as well as a bump around $L=350$~keVee due to double scattering off protons are indicated in the figure.}
\label{fig-light-resp}       
\end{figure}

Qualitatively, the light output recorded from neutron interactions in the detector medium allows us to determine a lower limit for kinetic energy of the incident neutron. However, due to the nonlinear conversion of recoil energy into scintillation light and the stochastic nature of the scattering process, there is a large variation in light output from the scintillator for events with a given incident neutron energy. A quantitative estimate of the probability that a neutron of a given kinetic energy produces a certain light output is given by the light-output response function $R(L|E_n)$. We have calculated it using the Monte-Carlo code Geant4~\cite{Geant4}, in order to account for the possibility of several scattering events inside the detector by the same neutron as well as other types of interactions. In the Monte-Carlo code we use the proton light-output function from Enqvist et al.~\cite{Enqvist2013} and also take the detector energy resolution into account. An example of the light-output response function is shown in Fig.~\ref{fig-light-resp}, where it is compared with experimental data from Enqvist et al. \cite{Enqvist2013}. This information is then used when calculating the PDF for each event.

Given an event $E$, the probability of the hypothesis $H_i$ to be true is then calculated as
\begin{equation}
P(E | H_i) = \Phi(E_n) \, R(L|E_n)\delta E_n,
\end{equation}
where $\Phi(E_n)$ is the prompt fission neutron spectrum, $\delta E_n$ is the neutron kinetic energy resolution and the index $i$ refers to a voxel. 
%
%
The total probability that the event occurs is
\begin{equation}
P(E) = \sum_i{P(E | H_i)}.
\end{equation}
The probability that the hypothesis is true is initially unknown and in fact the quantity that we are looking for. As initial guess it is taken to be uniformly distributed in the space defined by the voxels
\begin{equation}
P_0(H_i) = 1/n, \quad \forall i,
\end{equation}
where $n$ is the number of voxels. The probability that the hypothesis is true is updated after each event. If we would be sure that all events that we observe are from the source that we are looking for, it would be natural to use
\begin{equation}
P_k(H_i) = P_{k-1}(H_i | E),
\end{equation}
where $k$ indexes the event number. In this case the convergence is very fast, however, any spurious event due to accidental coincidence with background or scattered neutrons have a detrimental effect on the final result. In order to make the convergence more reliable we update the image according to
\begin{align}
P_k(H_i) &= (1-w){P_{k-1}(H_i) + w P_{k-1}(H_i | E)}, \\
    w &= 
\begin{cases}
    \alpha P_{k-1}(E),& \text{if } \alpha P_{k-1}(E)< 1\\
    1,              & \text{otherwise}
\end{cases}
\end{align}
where $\alpha$ is a weighting factor which is chosen to have a good compromise between fast convergence and low sensitivity to spurious events. In the present study a value of $\alpha=10$ was chosen.  Note the use of $P_{k-1}(E)$ in the weighted sum, this quantity is small for events whose probability density function have a small overlap with the current guess for the source location probability. Thereby it gives a way to suppress events that do not agree with the majority of events.

The result of the algorithm is the PDF of finding the source as a function of the position. The application of the algorithm to experimental data from the RPM with the source in different positions is illustrated in Fig. ~\ref{fig:Bayes_localization}. After 10 s of measurement with the geometry and source used in the experiments (1900 fissions/sec), the source location PDF has a standard deviation of about 4 cm. The absolute accuracy of the most probable location for long measurement times with the present detector geometry is given by a point spread function with an average standard deviation of about 1 cm.

\subsubsection*{Imaging multiple and distributed sources}
\begin{figure}[ht]
\centering
\includegraphics[width=0.6\columnwidth,clip]{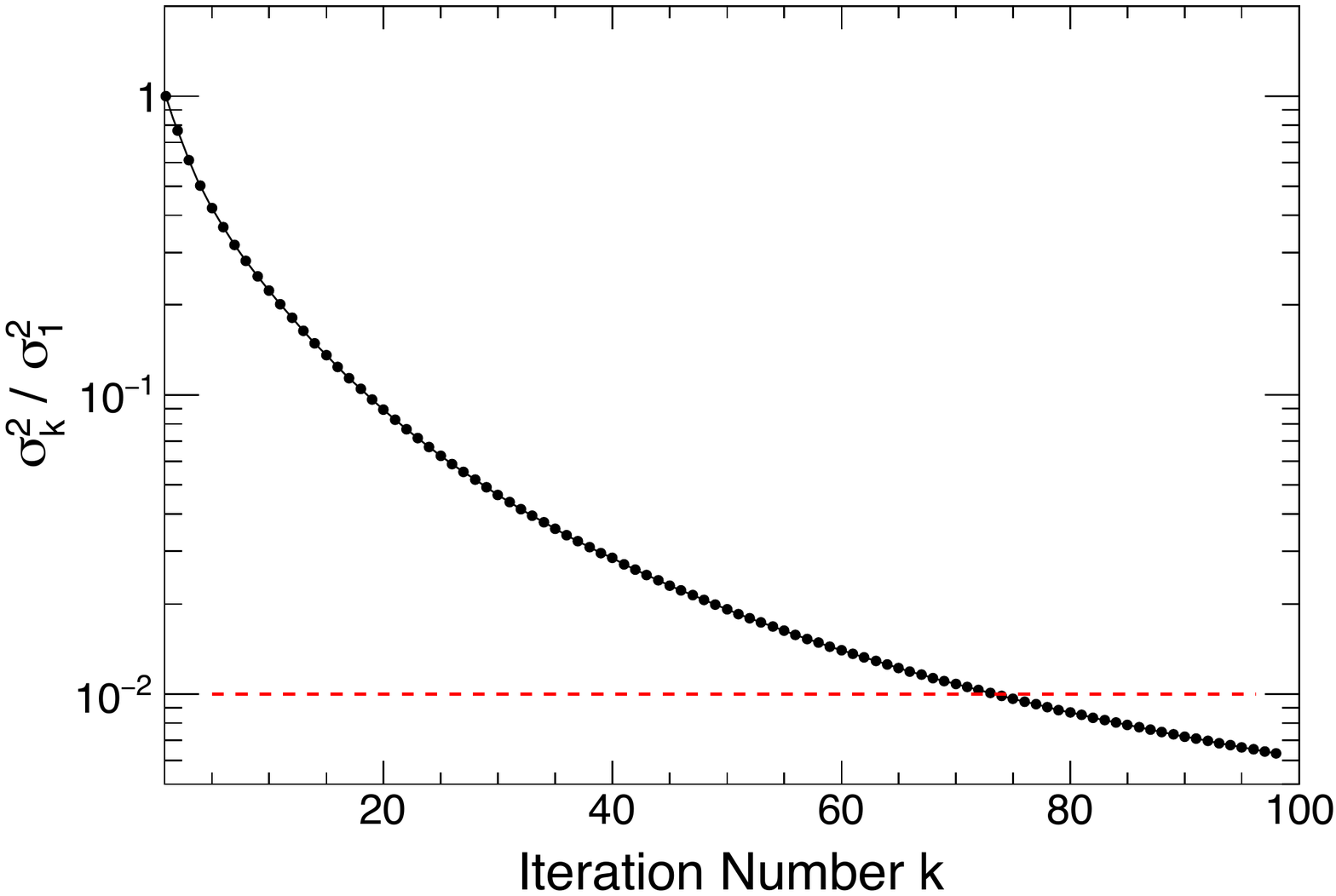}
\caption{Relative variance in image difference as a function of the iteration number, obtained from the data presented in Fig. 5 (Demonstration of imaging capability). The convergence criterion of 1 \% relative image difference is indicated with a red dashed line, and is achieved after 75 iterations.}
\label{fig:convergence}       
\end{figure}
The basic assumption made in the source localization algorithm is that all events originate from the same source position. If more than one source is present the algorithm will, at best, find only the most intense source. However, image reconstruction can be applied in a similar way to the Bayesian source localization algorithm. Instead of treating each event separately, accumulated data on $\Delta t_{n \gamma}$ and $L$ is used. From the accumulated data an image is de-convoluted using the \emph{iterative Bayesian unfolding} algorithm of D'Agostini \cite{DAgostini1994}. The number of coincidences originating from the voxel $i$ is estimated as
\begin{equation} \label{eq:unfolding}
n_C(H_i) \approx \sum_{j=1}^{n_E}P(H_i|E_j), 
\end{equation}
where $n_E$ is the number of events. The response function $P(E|H_i)$ is the same as in the source localisation algorithm described above. The de-convolution algorithm requires a first guess of the image $P(H)$ as starting point. The closer the initial guess is to the true image the better the result will be. In Fig.~\ref{fig:Bayes_localization_two_sources} we demonstrate the functionality of the method in the worst case scenario, i.e. no prior knowledge of the image.
As starting point, the prior is then taken to be uniformly distributed in space. The image intensity resulting from Eq.~(\ref{eq:unfolding}) is used as the prior in the next iteration.

When applying an iterative image de-convolution it is necessary to use a convergence criterion to determine when to stop the iteration. We use here the same approach as Steinberger et al. \cite{Steinberger2020}. The variance in image difference from two consecutive iterations is used to characterize how non-uniformly the image changed
\begin{equation}
\sigma^{2}_k = \text{Var}(\{ n_{C}(H_i)\}_k - \{ n_{C}(H_i)\}_{k+1}),
\end{equation}
where $\{ n_{C}(H_i)\}_k$ is the vector of voxel intensities for iteration number $k$. In Fig. \ref{fig:convergence} this parameter is plotted as a function of the iteration number, for the data presented in Fig. \ref{fig:Bayes_localization_two_sources}. We have chosen the convergence criterion $\sigma^{2}_k \leq 1 \%$.

\subsection*{NGET imaging based on cumulative time difference distributions (CNGET)}
For a system of $N$ detector elements there are $N(N-1)$ unique time difference distributions, taking into account which detector element detected the first in a time-correlated pair of particles emitted from a fission event. The CNGET technique uses this complete set of cumulative time difference distributions to determine the location of SNM in 3D using machine learning. With this technique one may, similarly to event-by-event-based NGET, analyze these time difference distributions continuously as they are updated during a measurement, successively improving on the accuracy of the localization.
Although not discussed in the present work, the method can also be applied to imaging of multiple or distributed sources using iterative image reconstruction methods.

\begin{figure}[h!]
\centering
\includegraphics[width=0.6\columnwidth,clip]{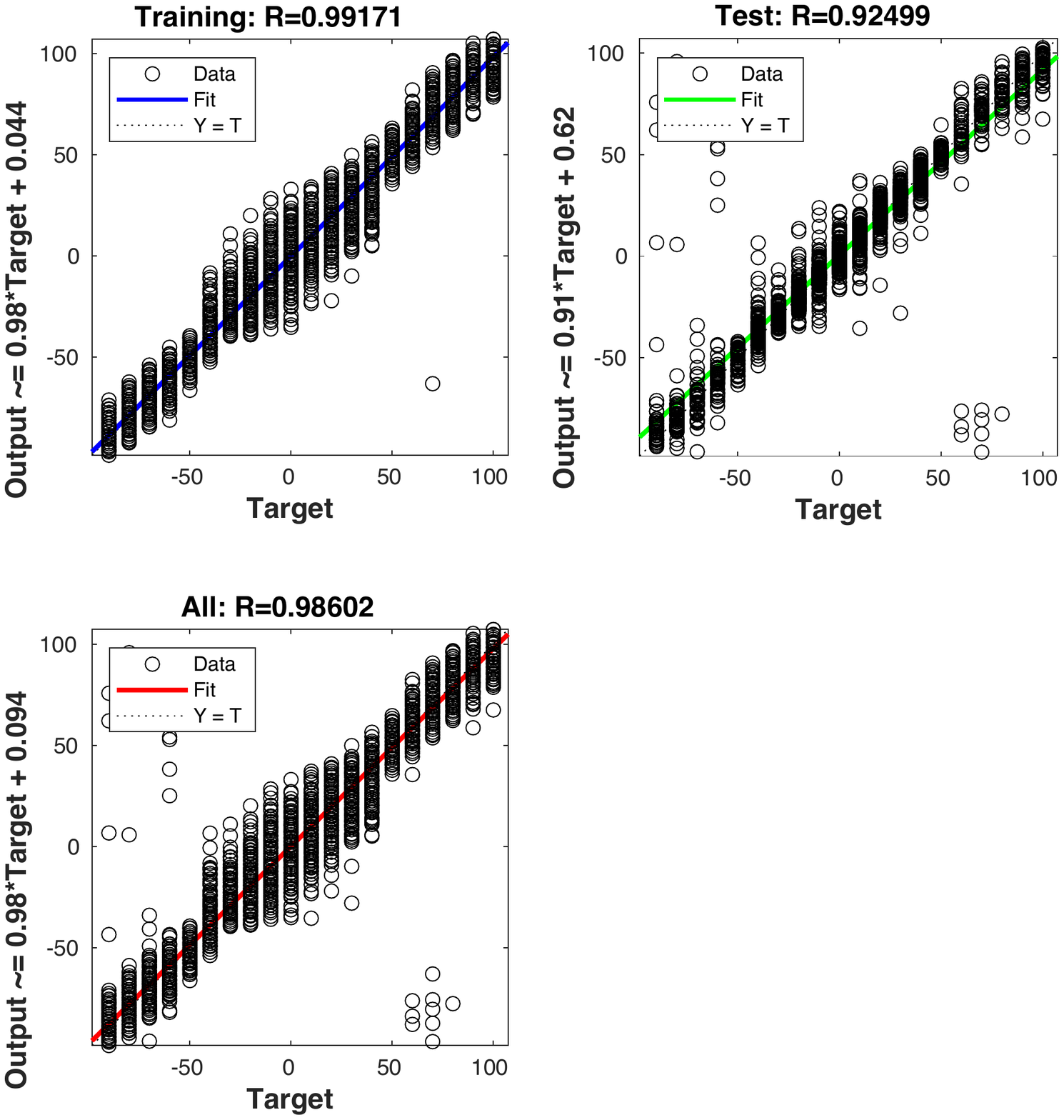}
\caption{Network output values as a function of target values for training and test data sets.The R value shows the relationship between the target and output values. The blue, green and red lines indicate the best linear regression fit for these data. }
\label{fig:ANN_regression}       
\end{figure}

Machine learning based on a four-layer feed-forward ANN was used to analyze the data in the present work.
The Neural Network Fitting App (nftool) from the Deep Learning Toolbox™ in the MATLAB$^{TM}$ computational software is used for training and testing. Nftool can solve the data-fitting problems using a two layer feed-forward network. From this app, a MATLAB script is generated to modify the training process.
The training data contain the matrix of input vectors as columns. Each vector includes the first, second, third and fourth moment extracted from the $\gamma$-neutron time difference distributions, for all 56 detector combinations. The target vectors are the actual positions of the Cf-252 source for each measurement. During the training process, the training and test data are randomly divided into two sets, so that 90\% of the data is used for training and 10\% is used as a completely independent test of network generalization.
The ANN employed a sigmoid transfer function in the three hidden layers and a linear transfer function in the output layer. The number of hidden neurons was set to ten in each of the three hidden layers. The best prediction results were achieved when the network was trained using a Bayesian Regularization algorithm. During the training process, the performance of the network was monitored. The regression plot, displayed in Fig.~\ref{fig:ANN_regression}, displays the network outputs for training and test data. If the network outputs would be equal to the targets, the data would lie along a 45 degree line for the perfect fit and the R value would be equal to 1.0. For our problem, the fit was objectively good, with R values higher than 0.92 for all sets. The R value for the test set will after some training time reach its highest value and slowly start to decrease due to over-fitting. This is the sign to stop the training process.

\clearpage

\end{document}